\documentstyle[preprint,aps]{revtex}
\tightenlines
\begin{document}
\title{A simple analytical calculation of mean-field potential in
Heavy Nuclei}
\author{Mofazzal Azam and Rajesh S. Gowda}
\address{Bhabha Atomic Research Centre,
Mumbai, India}
\maketitle
\vskip .8 in
\begin{abstract}
In the many-body theory of particles with long range interaction
potential, such as the potential of Newtonian gravity and its
relativistic generalization, Newton-Birkhoff theorem plays a
very important role. This theorem is violated in the case of short
range interaction potential, such as the Yukawa potential in
nuclear physics. We discuss the relevance of this in the many-body
treatment of heavy nuclei. In particular, using techniques similar
to those in Newtonian gravity,
we calculate analytically the mean-field
potential in heavy nuclei. The mean-field potential thus calculated
depends on well known quantities such as inverse pion mass, pion-nuleon
coupling constant, nuclear radius and density.
The qualitative features of
the mean-field potential thus obtained is similiar to the well known
phenomenological Wood-Saxon potential.
%The central depth of the potential agrees well even quantitavely.

\end{abstract}
\newpage
In this paper, using simple analytical calculations, we intend
to bring out some features of motion of nucleons in a heavy
nucleus \cite{coh,des}.
Our calculations are based on two phenomenologically well known
facts: The density of nucleons in heavy nuclei is almost constant
and the inter-particle interactions in nuclei is given by the Yukawa
potential. These assumptions seem to be sufficient for the qualitative
understanding of the motion of nucleons in heavy nuclei.
The central ingredient in our calculations is that the Yukawa
potential violates Newton-Birkhoof theorem. In order to explain the
role played by the violation of this theorem by the Yukawa potential
in nuclear physics, we first formulate and proof the theorem in Newtonian
gravity. The proof reveals why such a theorem can not be valid for Yukawa
potential. In the many-body treatment of heavy nuclei with
inter-particle Yukawa interactions, we consider the nucleon density
to be constant. In order to justify this assumption,
in the following section, we argue why a heavy nucleus can
be treated as system of constant density. We show that the constant density
has more to do with the Pauli's exclusion principle and short range
repulsive core of the nucleons rather than any ergodicity or thermodynamic
equillibrium \cite{bru,bru1,bethe,bru2,caur}.
In the next two sections, we derive the mean field
potential and indicate how the surface tension of heavy nuclei
emerges from the violation of Newton-Birkhoff theorem.
The mean field potential thus obtained qualitatively agrees with the
phenomenological Wood-Saxon potential\cite{wood}.
%The central depth of potential agrees well even quantitaively.
At the end, we make some concluding remarks.
%\vskip .8 in
%\newpage

\section{The Newton-Birkhoff Theorem}
Newton-Birkhoff theorem in the theory of gravitation is a very
important theorem. The violation of this theorem by Yukawa potential is the
central ingredient in our calculation of the surface tension and the
mean field of heavy nuclei. Therefore, in this paper we provide a discussion
of this theorem in Newtonian gravity in detail. Let us consider a
spherically symmetric ball of mass $M$, mass density $\rho$ and radius $R$.\\
Theorem (Part-I): The potential due to the spherical ball at a distance
$r$, where $r~>R$ is the same as the potential due to point particle
of mass $M$ placed at the centre of spherical ball.\\
Let us consider a spherical shell of outer radius $R$ and inner radius $b$,
having constant mass density $\rho$.\\
Theorem (Part-II): The gravitational potential due the spherical
shell at any point $r$, where $r~<b$ is constant. Therefore, the force
at the point $r$ is zero.\\
This theorem also holds in general theory of relativity and is called
the Newton-Birkhoff theorem. Bellow we provide proof of the theorem
in the contex of Newtonian Gravity. The proof reveals why the
theorem is so special to Newtonian gravitatinal potential and does not
hold for other potentials.\\

We want to calculate the gravitational potential
and potential energy of a spherically
symmetric ball of radius $R$ and constant density $\rho$.The two-body
Newtonian gravitational potential energy is given by,
\begin{eqnarray}
U_N=-\frac{Gm_1m_2}{|\vec{r}-\vec{r'}|}
\end{eqnarray}
where $\vec{r}$ and $\vec{r'}$  are the locations of the point masses $m_1$
and $m_2$ respectively.
The Newtonian gravitational potential at a point $\vec{r}$ due to a point
mass $m$ at point $\vec{r'}$ is given by
\begin{eqnarray}
\Phi_N(\vec{r})=-\frac{Gm}{|\vec{r}-\vec{r'}|}
\end{eqnarray}

For a mass distribution with density $\rho(\vec{r'})$, the formulas above take
the following forms,
\begin{eqnarray}
\Phi_N(\vec{r})=-G\int\frac{\rho(\vec{r'})d^3\vec{r'}}{|\vec{r}-\vec{r'}|}
\end{eqnarray}
\begin{eqnarray}
U_N=-G\int\int\frac{\rho(\vec{r})\rho(\vec{r'})d^3\vec{r}
d^3\vec{r'}}{|\vec{r}-\vec{r'}|}
=\int\Phi_N(\vec{r})\rho(\vec{r})d^3\vec{r}
\end{eqnarray}

In spherical co-ordinate system the formula for $\Phi_{N}(r)$
gives,
\begin{eqnarray}
\Phi_{N}(r)=-G\rho\int\int\int\frac{r'^2dr'sin\theta d\theta d\phi}
{(r^2+r'^2-2rr'cos\theta)^{1/2}}
\end{eqnarray}
In the expression above, the integral over $\phi$ ranges from $0$ to
$2\pi$ and for $\theta$, it ranges from $0$ to $\pi$. The integral over
$\phi$ gives a factor of $2\pi$.
The integral over $\theta$ can be carried out by making the following change
of variable
\begin{eqnarray}
\xi^2=r^2+r'^2-2rr'cos\theta ;~\xi d\xi=rr'sin\theta d\theta~; \nonumber
\end{eqnarray}

The integration over $~\xi~$ ranges from
$~~\xi=(|r-r'|)~ ~to ~\xi=(r+r')~~ $
After carrying out integrations over $\phi$ and $\theta$, we obtain
\begin{eqnarray}
\Phi_N(r)=~-2\pi G\rho\int[(r+r')-|r-r'|]\frac{r'}{r}dr'
\end{eqnarray}
For $r>~a~(> r')$, and $0\leq r'\leq a$
\begin{eqnarray}
\Phi_N(r)=~-2\pi G\rho\int_{0}^{a}[(r+r')-(r-r')]\frac{r'}{r}dr'=
-4\pi \frac{G\rho}{r}\int_{0}^{a} r'^2 dr'=-\frac{4\pi a^3\rho G}{3r}=
-\frac{GM}{r}
\end{eqnarray}
For $r~< b~(<r') $, and $b\leq r'\leq R$
\begin{eqnarray}
\Phi_N(r)=~-2\pi G\rho\int_{0}^{a}[(r+r')-(r'-r)]\frac{r'}{r}dr'=
-4\pi G\rho\int_{b}^{R} r'dr'=-2\pi G\rho(R^2-b^2)
\end{eqnarray}
which is constant independent of $r~$.
Note that, in the equation
above, the cancellation of $r$ from the expression is due to the special
form of the Newtonian potential.
Since force is given by the gradient of the potential with respect to
$\vec{r}$,
$\vec{F}=-\vec{\nabla}_{\vec{r}}\Phi_N(r)=0$, therefore
%grad_{\vec{r}}~[\Phi_N (r)]=0$,
there is no gravitational force within the spherical shell.
%Once we know the potential , it is easy to calculate the potential energy.
%It is given by,
%\begin{eqnarray}
%U_N=-G\int\int\frac{\rho(\vec{r})\rho(\vec{r'})d^3\vec{r}
%d^3\vec{r'}}{|\vec{r}-\vec{r'}|}
%=\int\Phi_N(\vec{r})\rho(\vec{r})d^3\vec{r}                \nonumber
%\end{eqnarray}
%Let us consider the case when $r>r'~$. Taking
%$a=r-\epsilon$, we obtain
%\begin{eqnarray}
%\Phi_N(r)=-\frac{4\pi G\rho (r-\epsilon)^3}{3r}    \nonumber
%\end{eqnarray}
%The potential energy is given by,
%\begin{eqnarray}
%U_N= =\int_{(|\vec{r}|\leq R)} \Phi_N(\vec{r})\rho(\vec{r})d^3\vec{r} \nonumber
%\end{eqnarray}
%Substituting the expression for $\Phi_N(r)$, we obtain,
%\begin{eqnarray}
%U_N =-\int_{\epsilon}^{R} \frac{4\pi
%G\rho (r-\epsilon)^3}{3r}\times 4\pi\rho r^2 dr    \nonumber
%\end{eqnarray}
%This integral above is convergent, and therefore, we can take the limit
%$\epsilon\rightarrow 0~$ before evaluating the integral.After integrating
%the resulting expression, we obtain
%\begin{eqnarray}
%U_N =--\frac{16\pi^2G\rho^2 R^5}{15}=-\frac{3}{5}\times \frac{GM^2}{R}
%\end{eqnarray}
This demostrates that in a classical theory
this constant potential is dynamically
irrelevant. However, it should be noted that in a quantum theory
this "constant potential well" is
relevant  and its' physical effects can be observed.

\section{Constant Density of Nucleons in Heavy Nuclei}
In this section we discuss why the density of nucleons in heavy nuclei
can be taken to be constant. We do not provide here a formal proof
of our claim. We rather argue that it follows very natuarally from
the short range character of the attractive potential, Pauli's exclusion
principle and the repulsive hard core potential of nucleons. Our claim
is essentially based on self consistency arguments.
However, it is important to note that
we do not require any ergodicity or thermodynamic equillibrium for
the constancy of the density of nucleons in the nuclei.\\
In order to maintain simplicity  in our exposition and to
avoid complications, we consider only one type of nucleon
and assume that it does not carry any electric charge. The attactive
interaction potential between a pair of nucleons is given by the
short range Yukawa Potential,
\begin{eqnarray}
V=-g\frac{e^{-\mu r}}{r}
\end{eqnarray}
where $g$ is the nuclear charge
( $\frac{g^2}{c\hbar}$ is the dimensionless nuclear
fine structure constant ) and $\mu$ characterizes the range of the attractive
potential. We assume that $d$ is the radius of the core of
the hard core potential.
We assume that, $R$, is the radius of the nucleus and $N$
is the number nucleons in it. At first, we consider the nuclear
matter limit of the nucleus. This means that both $R$ and $N$ are
large. Yukawa potential is a short range potential,
and therefore, each nucleon interacts
with nucleons in a small neighbourhood around itself.
The number of nucleons in its neighbourhood is proportional
to the density of nucleons $n_{\rho}$ . Let the constant of proportionality
be $f$ . Note that $f$ is the volume charaterising the range of the
potential and can be taken to be $f\approx \frac{1}{\mu^3}$.
Let us represent by $-\sigma$ the potential energy of attractive interaction
between two nucleons. Therefore, the potential energy of attractive
interaction of a single nucleon with its neighbours
is $-f n_{\rho} \sigma $ , and of $N$ nucleons with their neighbours is
$-fN n_{\rho} \sigma $. In the next section, we will show that
$~f\sigma =4\pi g^2/\mu^2~$ (see the end of section-III).\\
We assume that the kinetic energy of the nucleons
in the nucleus is solely due to the zero point energy.
Therefore, using Pauli's
exclusion principle we obtain,
$K.E. \approx \frac{\hbar^2}{2m}\times \frac{N^{5/3}}{R^2}$.
Taking $n_{\rho}\approx \frac{N}{4\pi/3~R^3}$, we write the total energy
$E(R)$ as function of $R$ as,
\begin{eqnarray}
E(R)\approx \frac{\hbar^2}{2m}\times \frac{N^{5/3}}{R^2}-fN n_{\rho}\sigma
= \frac{\hbar^2}{2m}\times \frac{N^{5/3}}{R^2}-\frac{3N^2f\sigma }{4\pi R^3}
\end{eqnarray}
We look for the variational (QM) ground state of the system. This can
be done by finding the minimum of the function E(R).
By imposing the condition, $\frac{dE}{dR}=0$, we obtain
all the extrema.
It turns that the function has a minimum
at $R=0$ ($E(R=0)=-\infty$) and a maximum when the interparticle distance
($=\frac{R}{N^{1/3}}$) is a few fermi. The function
goes to zero at infinity. Therefore, there is no minima for finite value
of $R$. Such a system would either fly apart or collapse to a point.
The ground state of the system is unstable. It is here that the hard core
potential comes to play an important role. Because of the hard core potential
the collapse would stop when $\frac{R}{N^{1/3}}=d$ . Therefore,
\begin{eqnarray}
E(R)
= \frac{\hbar^2}{2m}\times \frac{N^{5/3}}{R^2}-\frac{3N^2f\sigma }{4\pi R^3}
= \frac{\hbar^2}{2m}\times \frac{N}{d^2}-\frac{3Nf\sigma }{4\pi d^3}
\end{eqnarray}
and the per particle binding energy
\begin{eqnarray}
E(R)/N ~
= \frac{\hbar^2}{2m}\times \frac{1}{d^2}-\frac{3f\sigma }{4\pi d^3}
\end{eqnarray}
This leads to the emergence of constant density.
This picture resembles very much like those in packed granular material.
However, our arguments are exact only in the limit of nuclear matter.
For finite nuclei, however heavy it may be, volume occupied by the
nucleons will be a bit larger than the hard core radii. Pauli's
exclusion principle adding a soft core in addition to the already
existing hard core. This soft core is not a scattering potential. Because
of Pauli's exclusion principle , the nucleons rather avoid each other.
It is a granular material in which the particle fail to touch other
by the virtue of Pauli's exclusion principle.
But the gap is limited and this does not allow the particles to pile up
at any one single place thereby mainintaining a constant density profile.

\section{Derivation of the mean field potential in heavy nuclei }

We consider heavy spherical nucleus consisting of $N$ nucleons
and of radius $R$ . The density of nucleons in the nucleus,
$n_{\rho}$, is considered to be constant.
Since the system is spherically symmetric, the potential at any point
on a layer at $r$ will be the same. In general the potential on
any layer at $r$ will consist of two parts- the potential due to
the spherical ball of radius $r-\epsilon$ and the potential due to
the spherical shell of inner radius $r+\epsilon$ and outer radius $R$.
We will calculate these two contributions separately and then add them up.
This amounts to calculating the potential, $\Phi_{1}$,
out side a spherical ball and the potential , $\Phi_{2}$,
inside a spherical shell exactly in the way we carried out
the calculations in Newtonian gravity. Let us write the general
expression for the potential, $\Phi$,
\begin{eqnarray}
\Phi(\vec{r})=-g\int\frac{n_{\rho}
e^{-\mu|\vec{r}-\vec{r'}|}d^3\vec{r'} }{|\vec{r}-\vec{r'}|}
=-gn_{\rho} \int\frac{ e^{-\mu|\vec{r}-\vec{r'}|}
r'^2dr' sin\theta d\theta d\phi}{|\vec{r}-\vec{r'}|}
\end{eqnarray}

In the expression above, the integral over $\phi$ ranges from $0$ to
$2\pi$ and for $\theta$, it ranges from $0$ to $\pi$. The integral over
$\phi$ gives a factor of $2\pi$.
The integral over $\theta$ can be carried out by making the following change
of variable
\begin{eqnarray}
\xi^2=r^2+r'^2-2rr'cos\theta ;~\xi d\xi=rr'sin\theta d\theta~; \nonumber
\end{eqnarray}

The integration over $~\xi~$ ranges from
$~~\xi=(|r-r'|)~ ~to ~\xi=(r+r')~~ $
After carrying out integrations over $\phi$ and $\theta$, we obtain
\begin{eqnarray}
\Phi(r)=-\frac{2\pi gn_{\rho}}{\mu r}\int dr'~r'[e^{-\mu|r-r'|}
-e^{-\mu(r+r')}]
\end{eqnarray}
For $r>~a~(> r')$, and $0\leq r'\leq a$
\begin{eqnarray}
\Phi_{1}(r,a;~r>a)=-\frac{2\pi gn_{\rho}}{\mu r}
\int_{0}^{a} dr'~r'[e^{-\mu(r-r')} -e^{-\mu(r+r')}]
\end{eqnarray}
After carrying out the integration, we obtain
\begin{eqnarray}
\Phi_{1}(r,a;~r>a)=-\frac{4\pi gn_{\rho}}{\mu^2}\frac{e^{-\mu r}}{r}
[a~cosh(\mu a)- \frac{1}{\mu}sinh(\mu a)]
\end{eqnarray}
This is the potential out side the spherical ball of radius $a$
at a distance $r$ from its' centre. Let
\begin{eqnarray}
A(a)=-\frac{4\pi gn_{\rho}}{\mu^2}
[a~cosh(\mu a)- \frac{1}{\mu}sinh(\mu a)] \nonumber
\end{eqnarray}
This allows us to write the potential outside the spherical ball
as
\begin{eqnarray}
\Phi_{1}(r,a;~r>a)= -A(a)~\frac{e^{-\mu r}}{r} \nonumber
\end{eqnarray}
which has the same form as the Yukawa potential except for the fact
that, now, the nuclear charge is $A(a)$ and not $g$.To obtain the same
potential as that of the spherical ball at distance $r$, $~r>~a$
we need to place at centre of the ball a point nucleus of
nuclear charge equal to $A(a)$. We can estimate the value of $A(a)$
by writing the series expansion of hyperbolic cosine and sine
functions in the expression of $A(a)$. This gives,
%\begin{eqnarray}
%A(a)=\frac{4\pi gn_{\rho}}{\mu^2}
%[a~cosh(\mu a)- \frac{1}{\mu}sinh(\mu a)]
%=\frac{4\pi gn_{\rho}}{\mu^2}
%\Big{[}\sum_{n=1}^{\infty}\frac{2n}{(2n+1)!} \mu^{2n}a^{2n+1}\Big{]}
%\end{eqnarray}
%After some algebraic manipulations, we can write it as
%\begin{eqnarray}
%A(a) = \frac{4\pi a^3 gn_{\rho}}{3}
%\Big{[}\sum_{n=1}^{\infty}3\times\frac{2n}{(2n+1)!} (a \mu)^{2n-2}\Big{]}
%\end{eqnarray}
%We write the expression above as,
\begin{eqnarray}
A(a) =Ng\Big{[}1+\frac{1}{10}(\mu a)^2+\frac{1}{280} (\mu a)^4+
\sum_{n=4}^{\infty}3\times\frac{2n}{(2n+1)!}
(a \mu)^{2n-2}\Big{]}
\end{eqnarray}
where $N=\frac{4\pi a^3 n_{\rho}}{3}$ is the number of nucleons
inside the spherical ball of radius $a$. When radius $a$ is large,
the quantity $A$ is much much larger than the total nucleon charge
$Ng$ inside the spherical ball.
Therefore, to obtain the same
potential as that of the spherical ball containing the amount of
nuclear charge $Ng$ (~at distance $r$, $~r>~a$ ~),
we need to place at centre of the ball a point nucleus of
nuclear charge equal to $A$ which is larger than $Ng$.
This clearly demostrates that the
part-I of the Newton-Birkhoff theorem is violated by the Yukawa
potential.However, in the limit of $\mu\rightarrow 0$, we recover
result similar to the case of gravity. \\
For $r~< b~(<r') $, and $b\leq r'\leq R$
\begin{eqnarray}
\Phi_{2}~(r,b~;r<b)=-\frac{2\pi gn_{\rho}}{\mu r}\int_{b}^{R} dr'~r'
[e^{-\mu(r'-r)} -e^{-\mu(r+r')}]
\end{eqnarray}
After carrying out the integration, we obtain
\begin{eqnarray}
\Phi_{2}~(r,b~;r<b)=-\frac{4\pi gn_{\rho}}{\mu^2}\frac{sinh(\mu r)}{r}
[-e^{-\mu R}(R+\frac{1}{\mu})+e^{-\mu b} (b+\frac{1}{\mu})]
\end{eqnarray}
$\Phi_{2}~(r,b~;r<b)$ is the potential inside a hollow  spherical shell,
$\vec{F}=-\vec{\nabla}_{\vec{r}}\Phi_{2}~(r,b~;r<b)\neq 0$, and therefore,
there is force inside a hollow spherical shell.
This violates part-II of the Newton-Birkhoff theorem.
However, in the limit of $\mu\rightarrow 0$, we recover
result similar to the case of gravity. \\

It is now easy to calculate the potential at any $r$ within the nucleus.
Assuming,
\begin{eqnarray}
B(x)=\frac{4\pi g n_{\rho}}{\mu^2}(x+\frac{1}{\mu})e^{-\mu x}
\end{eqnarray}
We can write the total potential at $r$, $a<r<b$ as
\begin{eqnarray}
\Phi_{in}(r) =-A(a)\frac{e^{-\mu r}}{r}-~B(b)\frac{sinh(\mu r)}{r}
+~B(R)~\frac{sinh(\mu r)}{r}
\end{eqnarray}
When the radius of the inner sphere, $~a~$,  approaches $~r~$ from
bellow and the inner radius of the shell, $~b~$, approaches $~r~$
from above , the first two terms of the equation above add up to
a constant independent of $~r~$, the third term remains unchanged
and we obtain
\begin{eqnarray}
\Phi_{in}(r)=-\frac{4\pi gn_{\rho}}{\mu^2}
\Big{[}~1-\Big{(}R+\frac{1}{\mu}\Big{)}~
e^{-\mu R}~\frac{sinh(\mu r)}{r}~\Big{]}  \nonumber
\end{eqnarray}

%At any point on layer at $r$,
%the potential due to the inner sphere of radius $r-\epsilon$ is obtained
%by replacing $a$ by $r-\epsilon$ in the formula for for
%$\Phi_{1}(r,a~;r>a)$. This gives
%\begin{eqnarray}
%\Phi_{1}(r;a=r-\epsilon)=-\frac{4\pi gn_{\rho}}{\mu^2}\frac{e^{-\mu r}}{r}
%[r~cosh(\mu r)- \frac{1}{\mu}sinh(\mu r)] +{\mathcal{O}}
%(\frac{\epsilon}{r})
%\end{eqnarray}
%Similarly, the potential at any layer at $r$, due to the outer shell
%of inner radius $b=r+\epsilon$ and outer radius $R$, is obtained
%simply by replacing $b$ by $r+\epsilon$  in the formula for
%$\Phi_{2}(r,b~;r<b)$. This gives,
%\begin{eqnarray}
%\Phi_{2}(r;b=r+\epsilon)=-\frac{4\pi gn_{\rho}}{\mu^2}\frac{sinh(\mu r)}{r}
%\Big{[}~-e^{-\mu R}(R+\frac{1}{\mu})+e^{-\mu r} (r+\frac{1}{\mu})~\Big{]}
%+{\mathcal{O}}(\frac{\epsilon}{r})
%\end{eqnarray}
%The total potential, $\Phi(r)$, is given by
%\begin{eqnarray}
%\Phi(r)=\Phi_{1}(r;a=r-\epsilon)+\Phi_{2}(r;b=r+\epsilon) \nonumber
%\end{eqnarray}
%Therefore,
%\begin{eqnarray}
%\Phi(r)= -\frac{4\pi gn_{\rho}}{\mu^2}
%\Big{[}~1-\frac{1}{2r}(R+\frac{1}{\mu}) (e^{\mu(r-R)}-e^{-\mu(r+R)})~\Big{]}
%+{\mathcal{O}}(\frac{\epsilon}{r}) \nonumber
%\end{eqnarray}
%This can also be written as,
%\begin{eqnarray}
%\Phi(r)=-\frac{4\pi gn_{\rho}}{\mu^2}
%\Big{[}~1-\Big{(}R+\frac{1}{\mu}\Big{)}~
%e^{-\mu R}~\frac{sinh(\mu r)}{r}~\Big{]}
%+{\mathcal{O}}(\frac{\epsilon}{r})
%\end{eqnarray}

This is the mean field potential of the nucleons at any point $r$
inside the nucleus.
The mean potential outside the nucleus is given by
\begin{eqnarray}
\Phi_{out}(r)=-\frac{4\pi gn_{\rho}}{\mu^2}\frac{e^{-\mu r}}{r}
\Big{[}~R~cosh(\mu R)- \frac{1}{\mu}sinh(\mu R)~\Big{]}
\end{eqnarray}

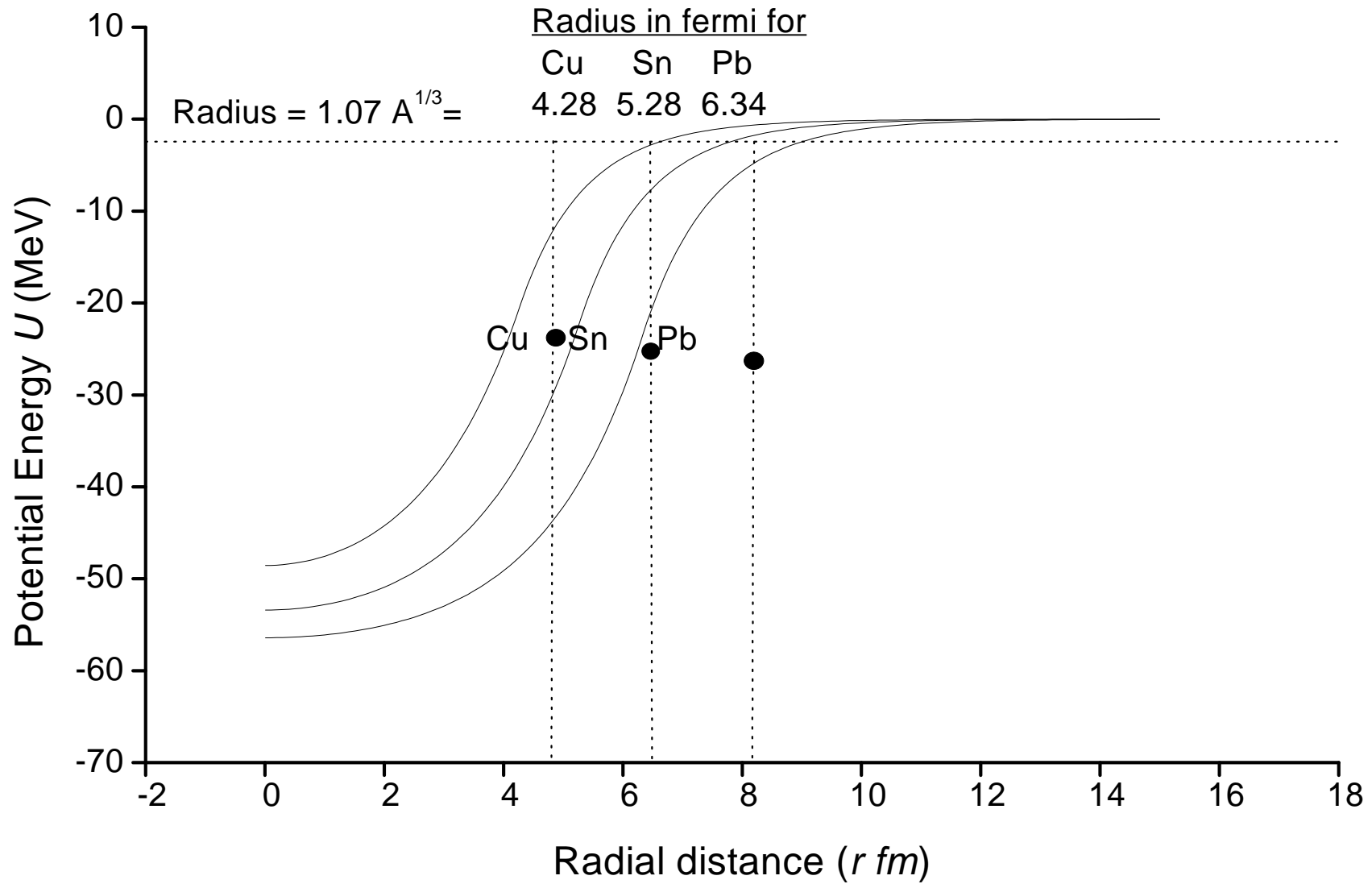
The potential energy  of a single nucleon in the mean mean field
inside and outside the nucleus is given by the following equations,
\begin{eqnarray}
U(r<R)=-\frac{4\pi g^2 n_{\rho}}{\mu^2}
\Big{[}~1-\Big{(}R+\frac{1}{\mu}\Big{)}~
e^{-\mu R}~\frac{sinh(\mu r)}{r}~\Big{]}
\end{eqnarray}
\begin{eqnarray}
U(r>R)=-\frac{4\pi g^2 n_{\rho}}{\mu^2}\frac{e^{-\mu r}}{r}
\Big{[}~R~cosh(\mu R)- \frac{1}{\mu}sinh(\mu R)~\Big{]}
\end{eqnarray}

%\begin{eqnarray}
%U(r<R)=-\frac{4\pi g^2n_{\rho}}{\mu^2}
%[1-\frac{1}{2r}(R+\frac{1}{\mu}) (e^{\mu(r-R)}-e^{-\mu(r+R)})]
%+{\mathcal{O}}(\frac{\epsilon}{r})
%\end{eqnarray}
%\begin{eqnarray}
%U(r>R)=-\frac{4\pi g^2n_{\rho}}{\mu^2}\frac{e^{-\mu r}}{r}
%[R~cosh(\mu R)- \frac{1}{\mu}sinh(\mu R)]
%\end{eqnarray}
In the formula above the values of the the physical constants
are the following:  pion-nucleon coupling constant,
$g^2/c\hbar =0.081$ , conversion constant, $ c\hbar=197.328$,
range of two-body Yukawa potential, $1/\mu=1.4094~fm$,
radius of nucleus of atomic number $A$, $~R=1.07~A^{1/3}~fm$,
density of nucleons in nucleus, $n_{\rho}=0.0165/fm^{3}$.
We provide plots of the potential for three different nuclei in
Fig.(1). The qualitative feature of
this potential is similiar to the phenomenological
Wood-Saxon potential \cite{wood}. The qualitative features is similiar
even near the boundary of the nucleus.
This is a bit of a surprising result because the nucleon density near
the boudary of the nucleus decreases drastically
where as in our derivation
we have assumed constant density through out the nucleus. This
suggests that the mean field potential is not very sensitive to
the distribution of nucleons near the boundary.\\
From eq.(23), it is now easy to infer that in the
nuclear matter limit, i.e., when $R$ is very large,
$~U(r)= -\frac{4\pi g^2 n_{\rho}}{\mu^2}~$. Therefore,
the quantity, $f\sigma$ introduced in section-II is equal to
$~4\pi g^2/\mu^2~$.\\

\section{The origin of surface tension in heavy nuclei}
In this section we sketch some arguments to show
that the surface tension in heavy
nuclei has its' orgin in the violation of Newton-Birkhoff
theorem. Actual calculation of the surface tension will require
the true density profile of the nucleons in the nucleus. However,
in the frame work of the present paper the density is assumed to be
constant, and therefore, we will not attempt to calculate the
surface tension. We rather point out that , even under the assumption
of constant density, one can observe the origin of surface tension. The
true density profile influences only the
numerical value of the surface tension.\\

To make the role played by the violation Newton-Birkhoff theorem
more transparent, we briefly recall some of
the derivations.
%Assuming,
%\begin{eqnarray}
%B(x)=\frac{4\pi g^2 n_{\rho}}{\mu^2}(x+\frac{1}{\mu})e^{-\mu x}
%\end{eqnarray}
%We can write the total potential at $r$, $a<r<b$ as
The total potential at  any point, $r$, in the region
between a sphere of radius  $a$ and a spherical shell of
inner radius $b$ and outer radius $R$ ($~~a~<r~<b~$)
is given by
\begin{eqnarray}
\Phi(r) =-A(a)\frac{e^{-\mu r}}{r}-~B(b)\frac{sinh(\mu r)}{r}
+~B(R)~\frac{sinh(\mu r)}{r} \nonumber
\end{eqnarray}
When the radius of the inner sphere, $~a~$,  approaches $~r~$ from
bellow and the inner radius of the shell, $~b~$, approaches $~r~$
from above , the first two terms of the equation above add up to
a constant independent of $~r~$, the third term remains unchanged
and we obtain
\begin{eqnarray}
\Phi_{in}(r)=-\frac{4\pi gn_{\rho}}{\mu^2}
\Big{[}~1-\Big{(}R+\frac{1}{\mu}\Big{)}~
e^{-\mu R}~\frac{sinh(\mu r)}{r}~\Big{]}  \nonumber
\end{eqnarray}
Force $\vec{F}$ , on a single nucleon at any point $r~$, $~r<R~$ is
given by,
\begin{eqnarray}
\vec{F}=-g\vec{\nabla}{\Phi_{in}(r)}=-\frac{4\pi g^2 n_{\rho}}{\mu^2}
~\Big{(}R+\frac{1}{\mu}\Big{)}~
e^{-\mu R}\Big{[} ~\frac{\mu r cosh(\mu r)-sinh(\mu r)}{r^2}~\Big{]}\vec{n}_r
\end{eqnarray}
and for $r~$,  $~r~>R$
\begin{eqnarray}
\vec{F}=-g\vec{\nabla}{\Phi_{out}(r)}=
-\frac{4\pi g^2n_{\rho}}{\mu^2}\frac{(1+\mu r)e^{-\mu r}}{r^2}
\Big{[}~R~cosh(\mu R)- \frac{1}{\mu}sinh(\mu R)~\Big{]}\vec{n}_r
\end{eqnarray}
where $\vec{n}_r$ is a unit vector directed along the radius towards
the centre of the nucleus.\\
The depth of the mean field potential varies very slowly
in the central region ($r\approx 0$) of the nucleus ( see Fig.(1) ).
However, near the outer surface, there is a sharp decline in the
depth of the total potential. Therefore, the force will have
very small in magnitude (practically zero)
in the central region but a broad peak near the bounday
of the nucleus. This is what we observe
in Fig.(2), where we have plotted the total force,
$\vec{F}$ as a function of $r$.
The direction of this  force at any point $r$ near the boundary
is towards the centre along the radius connecting the point.
Every nucleon in the outer shell is, therefore,
radially pulled towards the centre of the nucleus.
Since the force is practically zero in the central part of the
of the nucleus, there is a net pressure difference between the
outer shell and the inner core.
This is the origin of surface tension in nuclei. We believe
that the surface tension in most normal liquids have similar
origin . Our believe is based on the fact that potentials
of the type $-1/r^n$ which describe the properties liquid state
very well, can be obtained by the exchange of continuum
of Yukawa modes \cite{ran,azam} with some spectral weight $\rho(\mu)$.
In fact, the most commonly used Leonard-Jones potential in liquids
can be obtained with spectral weight $\rho(\mu)=\mu^4$.
\begin{eqnarray}
-\frac{1}{r^6}=-\frac{1}{4!}~\int_{0}^{\infty}~d\mu~ \mu^4~
\frac{e^{-\mu r}}{r}
\end{eqnarray}
\section{Concluding Remarks}
In this paper, we have tried to emphasise the importance of the
violation of Newton-Birkhoff theorem by Yukawa interaction for
the emergence of mean field shell model potential
as well as the surface tension in
heavy nuclei. The only input from the nuclear physics that we have
used is the constant density profile of nucleons in heavy nuclei.
Some very basic and general assumptions, such as
Pauli's exclusion principle, short range character of the attractive
Yukawa potential and the existence of a repulsive hard core in
nucleons, ensures the
emergence of constant density profile in heavy nuclei.
We have derived the mean field shell model
potential analytically. This potential qualitatively agrees wtih
the phenomenological Wood-Saxon potential.
Our formula explicitly shows how
the mean field potential
depends on pion-nucleon coupling constant, range
of the two-body Yukawa potential, nuclear size and the density.
Solution of the Schroedinger equation with the
mean field potential so derived leads to the emergence of
shell model structure. We have also pointed out how the
violation of the Newton-Birkhoff theorem
leads to the emergence of surface tension in heavy nuclei.
Our derivations, although semi-quantitative in nature,
strongly suggest that the qualitative features which
are attributed either to shell model or liquid drop model can be derived
from the Yukawa interaction of nucleons having repulsive hard core, and
some basic principles of quantum mechanics.

\section{Acknoweledgements}

The present work evolved from a conversation
with Prof. M.Sami  at the Inter-University Centre for Astronomy and
Astrophysics (IUCAA), Pune, India. We would like to thank him for
discussion, suggestions as well as hospitality at IUCAA.
M.A. would also like to thank Dr. S.G.Degweker of Bhabha Atomic
Research Centre, Mumbai, India, for critical comments and suggestions.
Thanks are also due to Dr. S.Ganesan of Bhabha Atomic Research Centre,
Mumbai, India, whose encouragement played an important role
in completion of this work.\\

\end{document}